# Constitutive Modeling of the Magneto-electromechanical Response of Composite Multiferroics Cylinders

Somer Nacy[1,2] and George Youssef[1,*]

[1]Experimental Mechanics Laboratory, Mechanical Engineering Department, San Diego State University, 5500 Campanile Drive, San Diego, CA 92182

[2]University of Baghdad, Jaderyia, Baghdad, Iraq

**Abstract**

Strain-mediated multiferroic composite structures are gaining scientific and technological attractions because of the promise of low power consumption and greater flexibility in material and geometry choices. In here, the direct magnetoelectric coupling coefficient (DME) of composite multiferroic cylinders, consisted of two mechanically bonded concentric cylinders, was analytically modeled under the influence of a radially emanating magnetic field. The analysis framework emphasized the effects of shear lag and demagnetization on the overall performance. The shear lag effect was analytically proven to have no bearing on the DME since it has no effect on the induced radial displacement due to the conditions imposed on the composite cylinder. The demagnetization effect was also thoroughly considered as a function of the imposed mechanical boundary conditions, geometrical dimensions of the composite cylinder, and the introduction of a thin elastic layer at the interface between the inner piezomagnetic and outer piezoelectric cylinders. The results indicate that the demagnetization effect adversely impacted the DME coefficient between 70% and 86%. In a trial to compensate for the reduction in peak DME coefficient due to the presence of demagnetization, non-dimensional geometrical analysis was carried out to identify the geometrical attributes corresponding to the maximum DME. It was observed that the peak DME coefficient is nearly unaffected by varying the inner radius of the composite cylinder, while it approaches its maximum value when the thickness of the piezoelectric cylinder is almost 60% of the total thickness of the composite cylinder. The latter conclusion was true for all of the considered boundary conditions.

**Corresponding Author:** gyoussef@sdsu.edu

**Nomenclature**

| | |
|---|---|
| $a$ | Inner radius of the inner cylinder |
| $b$ | Radius at the interface between cylinders |
| $c$ | Outer radius of the outer cylinder |
| $t$ | Thickness of the elastic layer |
| $r$ | Radial direction |
| $\theta$ | Hoop direction |
| $E$ | Modulus of elasticity of the elastic layer |
| $G$ | Modulus of rigidity |
| $\rho$ | Mass density |
| $v$ | Poisson's ratio of the elastic layer material |
| $K_s$ | Stiffness of the elastic layer |
| $H_o$ | Magnetic field |
| $H_{app}$ | Applied magnetic field |
| $H_{eff}$ | Effective magnetic field |
| $N_d$ | Demagnetization factor |
| $\mu_r$ | Relative permittivity of the piezomagnetic material |
| $e_{ij}$ | Piezoelectric coefficients |
| $q_{ij}$ | Piezomagnetic coefficients |
| $C_{ij}$ | Elastic coefficients |
| $\varepsilon_{33}$ | Dielectric coefficient |
| $U_\theta$ | Hoop and radial displacement |
| $U_r$ | Radial displacement |
| $D_r$ | Electric displacement in the radial direction |
| $E_r$ | Electric field in the radial direction |
| $\omega$ | Frequency |
| $\gamma$ | Shear strain |
| $\tau$ | Shear stress |
| $F_\theta$ | Body forces in the hoop direction |
| $F_r$ | Body forces in the radial direction |
| $\sigma_{\theta\theta}$ | hoop stresses |
| $\sigma_{rr}$ | Radial stresses |
| $\alpha$ | Direct Magnetoelectric Coefficient |

*Subscripts*

| | |
|---|---|
| $M$ | Piezomagnetic cylinder |
| $E$ | Piezoelectric cylinder |

# 1. Introduction

The supposition of complexity in describing concentric cylinder multiferroic composite structures for magnetoelectric coupling does not stem from merely the simple description of the kinematics, preferably from the plethora of interactions between three physics domains; namely mechanics, electrostatics, and magnetism [1]. Typically, strain-mediated multiferroic composite structures consist of two or more phases of piezoelectric and piezomagnetic materials bonded together in different configurations [2–4]. In such a case, the magnetoelectric coupling is said to be bidirectional, where the application of a magnetic field through the piezomagnetic material results in a spontaneous change in polarization within the piezoelectric phase through the transduction of strain across the interface. Converse coupling is also present, where an electric field applied across the piezoelectric material generates a mechanical strain that transduces at the interface yielding a change in the state of magnetization in the piezomagnetic material. It is, however, imperative to note that the cylindrical coordinate system, i.e., the case of ring or cylinder structure, gives rise to intricate physical interactions including self-boundedness, shape anisotropy, magnetic shielding, non-uniform strain distribution, and geometry and field-dependent magnetic states; to name a few [5,6]. Indeed, such coupled interactions are the motivation of describing the behavior of these structures as 'complex.'

With a focus on the interaction between the bias magnetic field and the continuum of the piezomagnetic materials, there are four direct and four converse magnetoelastic effects that delineated the strain-magnetization interdependency. In the direct sense, the geometry changes in the direction of the applied magnetic field in what is referred to as the Joule magnetostriction [1]. However, the application of a magnetic field also induces a change in the state of magnetization, which can result in change in the volume (i.e., volume magnetostriction) and a change in the elastic modulus (commonly denoted as the ΔE effect) [1,7]. When considering the converse coupling paradigm, there are inversely analogous effects, which are the Villari effect, the Nagaoka-Honda effect, and magnetically-induced changes in the elastic response, respectively [1]. Concurrent to these geometry-independent effects, there are three additional kinematically-induced effects; namely magnetic shielding, shape anisotropy, and onion state of magnetization [8]. The magnetic field preferentially permeates through the walls of a ferromagnetic cylinder due to the higher permeability than the surrounding air media resulting in shielding the air inclusion created by the walls of the cylinder. When applied diametrically, the magnetic field creates non-uniform state of magnetization in the cylinder, whereas the magnetic field wraps around the walls of the cylinder in two symmetric half circles (i.e., onion state of magnetization) yielding a strain gradation from the inner to the outer diameter [9–14]. Collectively, these bidirectional and spontaneous effects play a major role in the overall performance of strain-mediated multiferroic composites given that a large volume fraction of the structure is made of the piezomagnetic cylinder.

The quest to describe the full magneto-electro-mechanical response of strain-mediated multiferroic concentric cylinder structures has been evident in the recent literature from experimental [11,12,15–34], computational [14,35], and analytical approaches [36,37,45–51,38–40,40–44]. The outcomes of these research efforts further culminate the justification of describing the concentric ring structure as complex. For example, composite ring structures have been experimentally studied, which were consisting of an outer piezoelectric cylinder (PZT, lead zirconate titanate) and an inner piezomagnetic ring (Terfenol-D, alloy of iron, terbium and dysprosium) operating under the converse magnetoelectric coupling paradigm [9–13]. Experimentally, it also has been shown that the direction of polarization of the piezoelectric cylinder, the quality and method of interfacing the cylinders, the direction and magnitude of applied bias magnetic field, the frequency and magnitude of the electric field, and duration of loading symbiotically influence the overall response [9–13]. It is worth noting that Stampfli et al. and Youssef et al. recently reported the results of computational investigations of the same composite cylinder structure operating under the converse magnetoelectric coupling approach [14]. The computational results were found to be generally in good agreement with the experimental data but also provided quantitative and qualitative insights into the above-mentioned interactions and effects of the magnetic field passing through the cylinder structure [14]. Furthermore, Yakubov et al. and Pan et al. investigated tri-layer cylinders consisting of negative or positive magnetostrictive materials deposited on the interior and the exterior surfaces of a thin PZT cylinder using electroless process [30,52]. They reported the direct magnetoelectric response corresponding to axially and diametrically applied bias magnetic field consisting of superimposed DC and AC components [30,52]. Overall, the outcomes of the existing experimental reports in the literature align with the presumptions of Bichurin and Viehland that the concentric composite cylinder structures are worthwhile the investigation for magnetoelectric coupling applications [53]. It is important, however, to note that there are no experimental investigations, to the knowledge of the authors at the time of publication of this paper, of the direct magnetoelectric coupling of composite cylinders under the influence of radially emanating magnetic field given the practical challenges of experimentally replicating this situation, i.e., constructing a magnetic field source with radially emanating magnetic field; hence the persistence focus on analytically investigating this boundary-value problem within the realm of continuum mechanics.

Starting by the pioneer analytical work of Wang et al. using the effective medium theory to the recent reports by Youssef et al., these analytical models focus on mechanistically describing the dynamic magnetoelectric response of concentric cylinders [54–58]. Wang et al. assumed a directly and perfectly bonded piezoelectric and magnetostrictive cylinders and investigated the effect of four mechanical boundary conditions on the overall direct magnetoelectric response [57,58]. Youssef et al. recently published subsequent analytical investigation to supplement Wang's model through the consideration of an expanded set of mechanical boundary conditions. They also accounted for the inclusion of an elastic bonding layer, systematic investigation of the strain distribution given is the prime mediator between the applied magnetic field and resulting change in polarization, and exploration of the failure due to the generated mechanical stresses in all constituents [54–56]. While these models appear to be comprehensive, they make no attempt to investigate the shear lag and demagnetization effects. The latter negates the applied magnetic field causing an additional source of non-uniformity of the distribution of the magnetic field within the investigated structure, as shown later. The former, however, is at the root of potential non-uniform stress/strain distribution. It is then the focus of the analytical research leading to this paper to investigate these remaining effects and their influence on the overall magneto-electro-mechanical response of strain-mediated multiferroic composites. In all, the outcomes of the reported research are believed to be essential for the development of magnetoelectric devices based on the cylinder geometry that is tolerant to mechanical failure since demagnetization and shear lag effects may result in strain localization leading to damage. At the outset, an effort has been dedicated to identifying the geometrical parameters leading to maximize the direct magnetoelectric response of these composites.

## 2. Theory and Problem Formulation

The problem of a multiferroic composite cylinder structure (Figure 1), consisting of bonded piezoelectric/piezomagnetic cylinders, continues to be the consideration of this research, where the direct magnetoelectric effect (Joule effect) is investigated. It is worth noting that in contrary to some prior work, the active material cylinders are presumed to be assembled together using a passive elastic layer that is perfectly bonded to each of the cylinders at separate surfaces. The basic formulation is based on the linear piezoelectric and piezomagnetic constitutive relationships that have been reported a priori by Wang *et al.* and Youssef *et al.* but for the sake of completion [54–58], a brief introduction is included below since the derivation is required to substantiate the newly investigated shear lag and demagnetization effects.

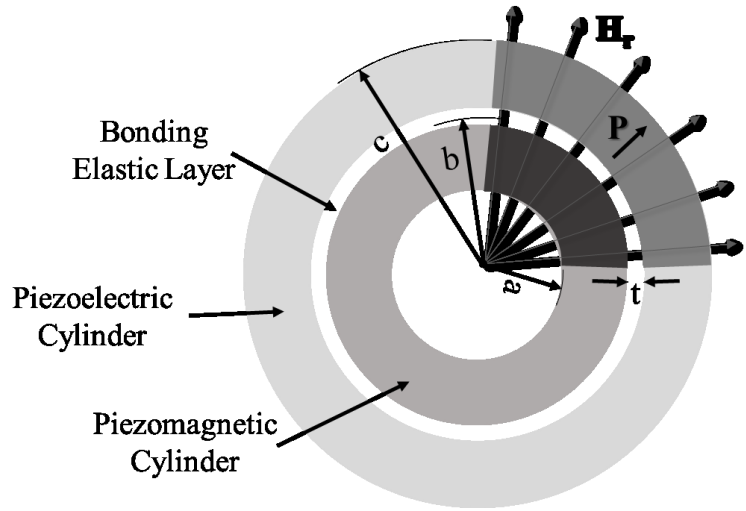

**Figure 1: Schematic of the geometry of the considered boundary-value problem.**

## 2.1 Basic Formulation

The outer cylinder is taken to be made of a piezoelectric material that is assumed to be radially polarized and mechanically orthotropic, while the inner cylinder is isotopic piezomagnetic under the effect of a time-harmonic uniform radially emanating magnetic field ($H_r = H_o e^{-i\omega t}$). The mechanics formulation is following plane strain assumption in the polar coordinate system (r, θ) per unit length of the cylinder. Therefore, the hoop and the radial stresses ($\sigma_{\theta\theta}$ and $\sigma_{rr}$, respectively) for the piezomagnetic cylinder can be written as

$$\sigma_{\theta\theta_M} = C_{11_M}\frac{U_{r_M}}{r} + C_{13_M}\frac{dU_{r_M}}{dr} - q_{13_M}H_o \qquad (1)$$

$$\sigma_{rr_M} = C_{31_M}\frac{U_{r_M}}{r} + C_{33_M}\frac{dU_{r_M}}{dr} - q_{33_M}H_o \qquad (2)$$

by noting: (1) the piezomagnetic material is electrically conductive such that $\{E\} = 0$; (2) the response due to the application of a magnetic field in one direction is independent from the magnetic response in the orthogonal directions (i.e., taken $B_\theta = 0$ and $H_\theta = 0$); and (3) the cylinder is actuating under a radially applied magnetic field $H_r = H_o$, where the time-harmonic factor ($e^{-i\omega t}$) was dropped to simplify the derivation given that all terms are similarly affected by it [57]. The Bessel differential equation for the piezomagnetic cylinder (Eqn. 3) is recovered after substituting the equations of stresses into the expression for mechanical equilibrium.

$$\frac{d^2U_{r_M}}{dr^2} + \frac{1}{r}\frac{dU_{r_M}}{dr} + \left(k_M^2 - \frac{\mu_M^2}{r^2}\right)U_{r_M} = \frac{QH_o}{r} \qquad (3)$$

where,

$$k_M^2 = \frac{\rho_M\omega^2}{C_{33_M}}, \qquad \mu_M^2 = \frac{C_{11_M}}{C_{33_M}}, \qquad \text{and} \qquad Q = \frac{q_{33_M} - q_{13_M}}{C_{33_M}}.$$

Following the same procedure for the outer piezoelectric cylinder, the components of the stresses in the polar coordinate system can also be written as shown in Eqn.4 and Eqn. 5.

$$\sigma_{\theta\theta_E} = C_{11_E}\frac{U_{r_E}}{r} + C_{13_E}\frac{dU_{r_E}}{dr} - e_{13_E}E_r \qquad (4)$$

$$\sigma_{rr_E} = C_{31_E}\frac{U_{r_E}}{r} + C_{33_E}\frac{dU_{r_E}}{dr} - e_{33_E}E_r \qquad (5)$$

In this case, three assumptions are applied to the outer cylinder based on the behavior of piezoelectric materials, which include: (1) the response of the piezoelectric cylinder is independent of the magnetic field; (2) charge accumulation at the outer surfaces of the cylinder is prohibited such that $D_r = 0$; and (3) the piezoelectric cylinder is considered to be radially polarized given rise to the condition of $E_\theta = 0$. Therefore, the equation of the radial electric displacement ($D_r$) can be written by following the linear piezoelectric constitutive relationship as shown in Eqn. 6.

$$D_{rE} = e_{31E}\frac{U_{rE}}{r} + e_{33E}\frac{dU_{rE}}{dr} + \epsilon_{33E}E_r \tag{6}$$

Again, a Bessel differential equation of the radial displacement in the piezoelectric cylinder as function of the properties is given by

$$\frac{d^2 U_{rE}}{dr^2} + \frac{1}{r}\frac{dU_{rE}}{dr} + \left(k_E^2 - \frac{\mu_E^2}{r^2}\right)U_{rE} = 0 \tag{7}$$

where,

$$k_E^2 = \frac{\rho_E \omega^2}{C_{oD}}, \qquad \mu_E^2 = \frac{C_{1D}}{C_{oD}},$$

$$C_{oD} = C_{33E} + e_{33E}e_{3DE}, \qquad C_{1D} = C_{11E} + e_{13E}e_{1DE}$$

$$e_{1D} = \frac{e_{31E}}{\epsilon_{33E}}, \quad \text{and} \quad e_{3D} = \frac{e_{33E}}{\epsilon_{33E}}.$$

As previously reported, the general solutions of the Bessel differential equation Eqn. 3 and Eqn. 7 are given in Eqn. 8 for the piezomagnetic cylinder and Eqn. 9 for the piezoelectric cylinder in terms of the first and second Bessel functions $J_\mu(\blacksquare)$ and $Y(\blacksquare)$ of order μ, respectively[57].

$$U_{rM}(r) = A_M J_{\mu_M}(k_M r) + B_M J_{-\mu_M}(k_M r) + H_o G(r) \tag{8}$$

$$U_{rE}(r) = A_E J_{\mu_E}(k_E r) + B_E J_{-\mu_E}(k_E r) \tag{9}$$

Where,

$$G(r) = Q\frac{\pi}{2}[Y_{\mu_M}(k_M r)\int_a^r J_{\mu_M}(k_M \zeta)d\zeta - J_{\mu_M}(k_M r)\int_a^r Y_{\mu_M}(k_M \zeta)d\zeta] \tag{10}$$

The unknown coefficients $A_E$, $B_E$, $A_M$ and $B_M$ in Eqn. 8 and Eqn. 9 are to be found from the boundary conditions and continuity conditions (discussed later).

**2.2 Shear lag Consideration**

New to this analytical modeling effort is the consideration of the shear lag effect of the bonding layer, whereas the terms associated with the shear stresses are now reintroduced into the expressions of mechanical equilibrium (Eqn. 11 and Eqn. 12).

$$\frac{1}{r}\frac{d\sigma_{\theta\theta}}{d\theta} + \frac{d\tau}{dr} + \frac{2\tau}{r} + F_\theta = 0 \tag{11}$$

$$\frac{d\sigma_{rr}}{dr} + \frac{1}{r}\frac{d\tau}{d\theta} + \frac{\sigma_{rr}-\sigma_{\theta\theta}}{r} + F_r = 0 \tag{12}$$

Since the problem at hand is axisymmetric and there are no boundary conditions imposed in the hoop direction, the hoop changes are nullified such that $\frac{d}{d\theta} = 0$ and the hoop body force is also zero ($F_\theta = 0$); therefore, the equilibrium equations can be reduced to Eqn. 13 and 14.

$$\frac{d\tau}{dr} + \frac{2\tau}{r} = 0 \tag{13}$$

$$\frac{d\sigma_{rr}}{dr} + \frac{\sigma_{rr}-\sigma_{\theta\theta}}{r} + F_r = 0 \tag{14}$$

Clearly, the solution of the differential equation shown in Eqn. 14 follows exactly the same procedure discussed before (and reported in detail later) since it is indeed identical to mechanical equilibrium expression leading to Eqn. 3 and 7 above. On the other hand, the consideration of the shear stress yields a hoop equilibrium condition (Eqn. 13), the solution of which can be obtained to be

$$\tau = Ar^{-2} \tag{15}$$

where, A is a constant to be found from the boundary conditions. However, the interface shear stress ($\tau_i$) induced at the interface is due to the hoop stresses ($\sigma_{\theta\theta}$) and, therefore, can be approximated by the interfacial hoop stresses in the piezomagnetic and piezoelectric cylinders. The interface shear stress is precisely the boundary condition necessary to calculate the unknown coefficient in Eqn. 15 and it follows that the shear stress distribution in the radial direction can also be obtained. Moreover, this induced shear stress is related to the shear strain ($\gamma$) and hence to the hoop displacement ($U_\theta$) as described in Eqn. 16.

$$\tau = G\gamma = G\left(\frac{dU_\theta}{dr} - \frac{U_\theta}{r}\right) \tag{16}$$

While it is obvious from the preceding discussion that the closed-form solution for the shear stress and hoop displacement can be easily obtained, there is a striking byproduct finding. The hoop-related parameters (i.e., shear stress and hoop displacement) have no effect on the expressions of stresses, strain, and displacement obtained in the basic formulation stated above. It is then reasonable to conclude that the shear lag effect has no influence on the sought after magnetoelectric response of concentric composite cylinders, given the assumptions leading to Eqn. 15 and Eqn. 16. In other words, the magnetoelectric response of concentric composite cylinder structure is immune from the shear lag effect as mathematically proven above, henceforth, the shear lag is not further considered for the remainder of the paper given that its inclusion would yield identical results to its absence since it has no effect on the radial displacement ($U_r$) obtained previously. However, it is important to note that this conclusion is only applicable in the context of the linear theory of elasticity, whereas material and response nonlinearities were suppressed. Future research will focus on exploring the effect of the nonlinear and time-dependent bonding layer on the overall response.

## 2.3 Demagnetization Effect Consideration

The passing of a magnetic field through a ferromagnetic material results in a change in magnetization, which in turn emanates a magnetic flux. However, the effective magnetic field is reduced from the applied magnetic field due to the demagnetization effect, where the latter is a function of the emanating magnetic flux from the sample that is corrected by a geometry-specific demagnetization factor ($N_d$). In all, the effective magnetic field ($H_{eff}$) acting on the sample can be expressed as shown in Eqn. 17, which is in turn used to substitute for the magnitude of the radial magnetic field ($H_o$) as discussed in Eqns. 1-3.

$$H_{eff} = H_{app}\left(\frac{1}{1+N_d(\mu_r-1)}\right) \quad (17)$$

The demagnetization factor has been documented before in the literature for many geometries and for single crystal and polycrystalline materials [8], however, the factor for a hollow cylinder has not yet been reported. To estimate the demagnetization factor without resorting to solving the Poisson's equation, the following assumptions are introduced, which stem from the kinematics of the concentric composite cylinder and the magnetic boundary condition. As discussed above, the magnetic field is applied radially outward from the center, hence $H_{app}$ is assumed to be applied uniformly over the circumference of the cylinder. That is to say, the demagnetization effect is occurring in a plane and in the radial direction only, where it is uniform circumferentially. This assumption then further simplifies the problem, where the cylindrical geometry is considered as a prism, as shown in Figure 2a. Hence according to Joseph et al. [59], (see Appendix A), the demagnetization factor is defined as

$$N_d = \frac{1}{4\pi}[2\cot^{-1}f(r\theta,r) + 2\cot^{-1}f(-r\theta,r) + 2\cot^{-1}f(r\theta,-r) + 2\cot^{-1}f(-r\theta,-r)]$$

(18)

where,

$$f(r\theta,r) = \frac{\left[(\pi r_m-r\theta)^2+\left(\frac{w}{2}\right)^2+\left(\frac{h}{2}-r\right)^2\right]^{\frac{1}{2}}\left(\frac{h}{2}-r\right)}{(\pi r_m-\theta r)\left(\frac{w}{2}\right)} \quad (19)$$

with $r_m$ is the mean radius, $h$ is the wall thickness and $w$ is the wall width, all dimensions are of the piezomagnetic cylinder. Since the magnetic field is considered to be uniformly distributed along the circumference, the contribution in the θ-direction is neglected. The demagnetization factor is then dependent on geometry and on the direction of the applied magnetic field. The demagnetization factor is plotted in Figure 2b as a function of the wall thickness ranging from 0.002 m to 0.02m. In the case under investigation here, the demagnetization factor is independent of the mean radius since the applied magnetic field was assumed to be uniform outward from the center. As thickness increases, the effect of the demagnetization field is minimized such that $N_d \approx 0$ indicating that the effective magnetic field becomes equivalent to the applied field. Generally, Figure 2b signifies the dependence of the demagnetization factor on the geometry of the piezomagnetic cylinder.

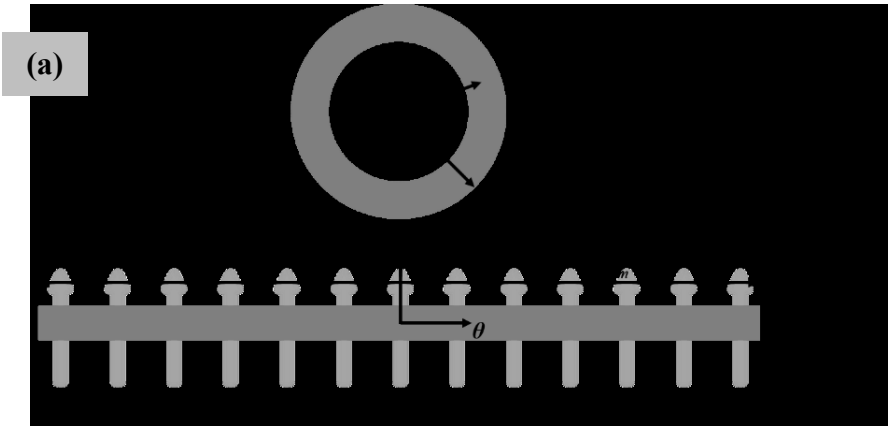

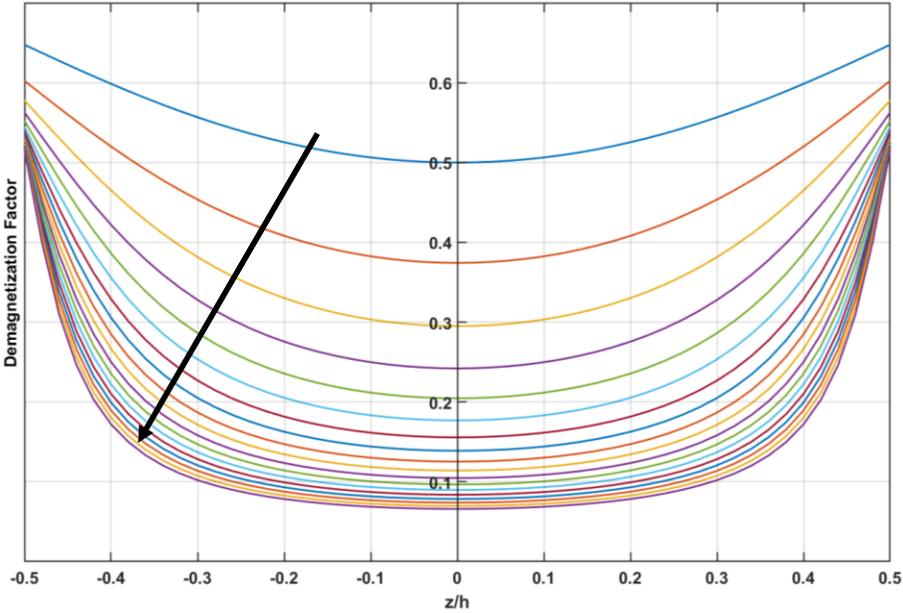

**Figure 2: (a) Schematic of (top) the inner piezomagnetic cylinder with its defining geometrical parameters and (bottom) prism resulting from unfolding the cylinder along its circumference, and (b) the calculated demagnetization factor as function thickness (arrow indicate increasing thickness from 0.002 to 0.02 m) based on Equation 18.**

## 2.4 Solution of the Boundary-value Problem (no demagnetization effect)

To find the unknown coefficients in Eqn. 8 and Eqn. 9, four mechanical boundary conditions are considered in addition to above-stated electrical and magnetic boundary conditions. The mechanical boundary conditions (Table 1) include free and clamped conditions on the most inner and outer surfaces of the composite cylinder.

**Table 1: List of considered mechanical boundary conditions at the inner ($r = a$) and outer ($r = c$) radii of the composite cylinder (Figure 1).**

| | Boundary Conditions | | | |
|---|---|---|---|---|
| Location | Free - Free | Clamped - Free | Free - Clamped | Clamped - Clamped |
| $r = a$ | $\sigma_{rr} = 0$ | $U_r = 0$ | $\sigma_{rr} = 0$ | $U_r = 0$ |
| $r = c$ | $\sigma_{rr} = 0$ | $\sigma_{rr} = 0$ | $U_r = 0$ | $U_r = 0$ |

The continuity boundary condition has been modified by Youssef et al. [56] after the inclusion of bonding elastic layer (Eqn. 20), where they discussed the effect of the elastic adhesive layer on the overall magnetoelectric coefficient.

$$\sigma_{rr_E}\left(b + \tfrac{t}{2}\right) = \sigma_{rr_M}\left(b - \tfrac{t}{2}\right) \quad \text{and} \quad \left[U_{r_E}\left(b + \tfrac{t}{2}\right) - U_{r_M}\left(b - \tfrac{t}{2}\right)\right] = \frac{1}{k_s}\left(2\pi b \sigma_{rr_M}\left(b - \tfrac{t}{2}\right)\right) \tag{20}$$

Where, $k_s$ is the characteristic stiffness of the bonding elastic layer and given by Eqn. 21 in terms of the elastic properties of the material (E is the modulus of elasticity and ν is Poisson's ratio) and the geometry of the bonding layer (r is the mean radius and t is the thickness).

$$k_s = \frac{2\pi r E(1-v)}{t(1+v)(1-2v)} \tag{21}$$

In the case of bonding layer-free interface where the piezoelectric and magnetic electric cylinders are directly bonded to one another, the continuity condition of direct bonding is shown in Eqn. 22.

$$\sigma_{rr_E}(b) = \sigma_{rr_M}(b) \quad \text{and} \quad \left[U_{r_E}(b) - U_{r_M}(b)\right] = 0 \tag{22}$$

Thereafter, the ME coupling coefficient (Eqn. 23) is calculated after applying the boundary and continuity conditions to find the unknown coefficients [56].

$$\alpha = \frac{e_{3D}[U_{r_E}(r=c) - U_{r_E}(r=b+t/2)] + e_{1D} \int_{(b+t/2)}^{c} r^{-1} U_{r_E} dr}{H_o(c-b+t/2)}. \quad (23)$$

**Table 2: Material Properties of the piezoelectric cylinder, the piezomagnetic cylinder, and the elastic bonding layer [35].**

| Material | Property | Value | Unit |
|---|---|---|---|
| PZT-5A | P | 7500 | [kg m$^{-3}$] |
| | $c_{11}$ | 99.201 | [GPa] |
| | $c_{13}$ | 50.778 | [GPa] |
| | $c_{33}$ | 86.856 | [GPa] |
| | $e_{13}$ | -7.209 | [N C$^{-1}$] |
| | $e_{33}$ | 15.118 | [N C$^{-1}$] |
| | $\varepsilon_{33}$ | 1.5 E-8 | [C$^2$ N$^{-1}$ m$^{-2}$] |
| Terfenol-D | P | 9200 | [kg m$^{-3}$] |
| | $c_{11}$ | 8.451 | [GPa] |
| | $c_{13}$ | 3.91 | [GPa] |
| | $c_{33}$ | 28.3 | [GPa] |
| | $q_{13}$ | -5.75 | [N A$^{-1}$ m$^{-1}$] |
| | $q_{33}$ | 270.1 | [N A$^{-1}$ m$^{-1}$] |
| Bonding Layer | E | 0.1 | [GPa] |
| | Y | 0.4 | |

**2.5 Solution of the Boundary-value Problem (with demagnetization effect)**

When the demagnetization effect is considered, the applied magnetic field ($H_o$) in Eqn. 1 is replaced by the effective magnetic field ($H_{eff}$) defined in Eqn. 17 as discussed above. This substitution implies that the forcing function is no longer a constant rather it becomes a function of the radial direction, i.e., H$_{eff}$ is a function of r since the demagnetization factor ($N_d$) is radially-dependent as shown in Eqn. 18. The resulting differential equation can then be solved by the Lagrange's method of variation of parameters, where the general solution consists of two terms $u_r = u_c + u_p$ such that $u_c$ is the complementary solution given by Eqn. 24 and $u_p$ is the particular solution defined by Eqn. 25.

$$u_c = AJ_{\mu M}(k_M r) + BY_{\mu M}(k_M r) \qquad (24)$$

$$u_p = J_{\mu M}(k_M r) \int_a^r \frac{H_{eff} Q Y_{\mu M}(k_M r)}{r\left[Y_{\mu M}(k_M r)J'_{\mu M}(k_M r) - Y'_{\mu M}(k_M r)J_{\mu M}(k_M r)\right]} dr +$$
$$Y_{\mu M}(k_M r) \int_a^r \frac{H_{eff} Q J_{\mu M}(k_M r)}{r\left[J_{\mu M}(k_M r)Y'_{\mu M}(k_M r) - J'_{\mu M}(k_M r)Y_{\mu M}(k_M r)\right]} dr \qquad (25)$$

Once the total solution is determined, the magnetoelectric coefficient can be calculated using the same procedure as above.

The material properties of the outer piezoelectric cylinder, elastic bonding layer, and inner piezomagnetic cylinder are listed in Table 2, which were used in obtaining the solution of the boundary-value problem with and without accounting for the demagnetization effect.

## 3. Results and Discussions

Figure 3 shows the direct magnetoelectric coupling coefficient as a function of frequency ranging from 0.1 MHz to 3 MHz for all four considered mechanical boundary conditions while elucidating the dependence of the DME on the presence of the elastic bonding layer and the demagnetization effect. The total radial thickness of the composite cylinder was 5 mm, whereas the inner radius was 10 mm, the interface radius was 12 mm, and the outer radius was 15 mm. The applied magnetic field was taken to be 60 kA/m. It is important to note that the previous analytical studies pointed towards the paramount importance of geometrical dimensions in preselecting the resonance frequency even more important than the values of the coupling coefficients [56]. When considered, the thickness of the elastic bonding layer was taken to be 7.5 µm based on the research outcomes from [56]. The composite Figure 3 also demonstrates the effect of the mechanical boundary conditions on the DME coupling coefficient. Specifically, the figure consists of four sub-figures, each representing clamped-clamped, free-free, free-clamped, and clamped-free boundary conditions defined on the inner and outer diameters of the composite cylinder, respectively. Also plotted at the bottom of each sub-figure is the difference between DME responses with and without the demagnetization factor accounted for when the bonding layer was suppressed or sanctioned. The difference subplots qualitatively and quantitatively show that the demagnetization effect is more pronounced in the response region corresponding to the initial dynamics, i.e., in the vicinity of the first and second harmonics. In all, the DME maxima and the associated resonant frequencies were found to be dependent on the boundary conditions, the demagnetization effect, and the bonding layer.

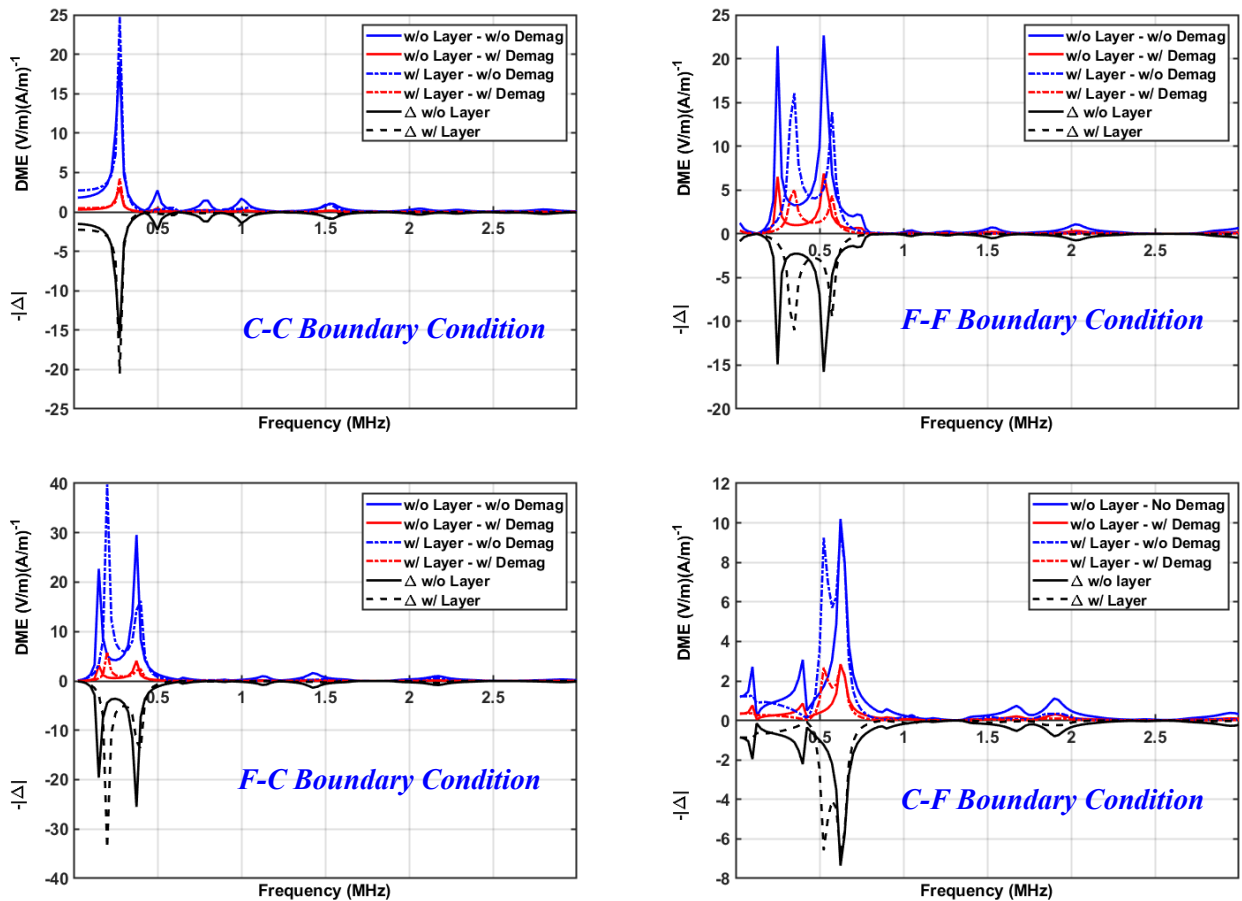

**Figure 3: The frequency-dependent DME response of concentric composite cylinder for four mechanical boudndary conditions at the bottom of each figure is the difference between the responses when the demagentization effect was considered and suppressed.**

As the frequency increased, beyond the fundamental harmonic, the resulting DME coupling coefficient is substantially lower, as expected, than the peak values at the resonance. For example, the DME values at the first resonant frequency for the F-F boundary condition was found to be 5.04 $(V/m)(A/m)^{-1}$ (at 350 kHz) and 6.51 $(V/m)(A/m)^{-1}$ (at 250 kHz) for the scenario when the demagnetization effect was considered in the presence and absence of the bonding layer, respectively. Thereafter, the demagnetization-modified peak DME value for the same boundary condition was merely 4.36 $(V/m)(A/m)^{-1}$ associated with a frequency of 575 kHz when the bonding layer was sanctioned. Correspondingly, the peak DME was 6.88 $(V/m)(A/m)^{-1}$, at 525 kHz in the absence of the bonding layer. The diminished high-frequency response is attributed to the deexcitation of radially-expanding vibrational modes, but this can be effectively tuned to meet specific design requirements through the manipulation of the geometrical and material attributes.

### *3.1 Effect of Boundary Conditions*

The mechanical boundary conditions play a major role in the overall DME response including the peak values, the corresponding frequencies, and the attributes of waveform of the resonant frequencies. In this section, the focus is on the interrelationship between the resulting DME coupling coefficient and the applied boundary condition, therefore, the discussion is limited to the case of absence of demagnetization and bonding layer (the remaining scenarios are considered in the following sections). The maximum DME response was found to be 29.57 $(V/m)(A/m)^{-1}$ when the inner diameter of the composite was mechanically free while the outer diameter was clamped at a frequency of 375 kHz. The maximum DME values for the remaining boundary conditions were extracted to be 19.06, 22.65, and 10.2 $(V/m)(A/m)^{-1}$, for the C-C, F-F, and C-F conditions, respectively, occurring at the corresponding frequencies of 275, and 525, and 625 kHz. It is worth noting that the electrical and magnetic boundary conditions remained unchanged throughout the analysis, as noted in the model section. The sensitivity of the magnetoelectric coupling coefficient to the change in the mechanical boundary conditions stems from two specific reasons.

First, the change in the boundary conditions correspond to a change in the apparent structural stiffness of the composite cylinder, hence not only shifting the frequency but also affecting the values of the coupling coefficient. In a recent research, our group formalized the interrelationship between a comprehensive set of mechanical boundary conditions, including those used herein, the materials properties, and the geometrical attributes of each of the constituents in what we termed the normalized stiffness parameter [55]. The latter was then used to investigate the sensitivity of the DME response signifying the contribution of each of the above-mentioned factors. The manipulation of the boundary conditions for the same geometrical and material attributes can dynamically tuning the DME frequency response for hardware-agnostic antennas and filters.

Second, the type of the applied boundary condition dedicates the distribution of the radial displacement field within each constituent phase of the composite. The results in Figure 3 clearly signify that the type (free vs. clamped) and location (inner vs. outer diameter) of a boundary condition have a defining contribution on the efficiency of strain transfer from the inner actuator piezomagnetic cylinder to the outer sensor piezoelectric cylinder. For example, clamping the outer surface of the piezoelectric cylinder while prescribing stress-free boundary condition at the inner diameter of the piezomagnetic cylinder (i.e., F-C) resulted in the highest DME coefficient since the transferred radial displacement to the outer cylinder resulted in an increase in electrical displacement due to the enhanced piezoelectric strain.

*3.2 Effect of Bonding Layer*

The addition of the bonding elastic layer has a profound effect on the magnetoelectric coupling coefficient as shown in Figure 3 while disregarding the demagnetization effect (discussed next). The DME values in the presence of a bonding layer were 29.6% and 34.5% higher than when the layer's effect was suppressed in the cases the outer diameter of the composite cylinder was clamped for the C-C and F-C conditions, respectively. On the other hand, i.e., when the outer diameter was free, the bonding layer appeared to have an adverse effect such that the DME values were 5.9% and 29.1% lower in comparison to the results when the bonding layer was absent for the C-F and F-F boundary conditions, respectively. For example, the peak values of the DME coefficient for the case of F-C boundary condition were found to be 39.8 $(V/m)(A/m)^{-1}$ and 29.6 $(V/m)(A/m)^{-1}$, when the elastic bonding layer was sanctioned and suppressed, respectively. Similarly, the maximum DME changed from 24.7 to 19.1, 9.6 to 10.2, and 16.1 to 22.7 $(V/m)(A/m)^{-1}$ for the C-C, C-F, and F-F conditions, respectively. The presence of the elastic layer and clamping the outer diameter promoted the transfer of the radial displacement from the actuator inner cylinder to the outer piezoelectric cylinder resulting in the improved DME values. In other words, the efficacy of the strain mediation was enhanced by the presence of the bonding layer that acted as a mechanical mediator transitioning the difference in the elastic properties of the inner and outer cylinders. On the other hand, constraining the outer boundary gave rise to higher piezoelectric coupling since the difference in the radial displacement across the piezoelectric cylinder is amplified. While a single thickness of the bonding layer was considered herein, the increase in thickness was recently reported to affect the underlying strain transduction phenomena that is primarily responsible for the magnetoelectric coupling paradigm under consideration [54].

In addition to affecting the amplitude of the DME coupling coefficient, accounting for the effect of an ultrathin bonding layer resulted in shifting the resonant frequencies. For example, the frequency of the first-harmonic was found to be 250, 150, and 350 kHz for F-F, F-C, and C-F boundary conditions, respectively, in the absence of the effect of the elastic layer. In contrast, these frequencies accordingly shifted to 350, 200, and 525 kHz due to modifying the continuity condition to account for the presence of the elastic layer. In the case of clamped-clamped mechanical boundary condition, and regardless of the inclusion or exclusion of the bonding layer effect, the frequency of the first harmonic remained unchanged at 275 kHz. The insensitivity of the resonant frequency in C-C boundary condition is attributed to the dominance of the stiffness of the piezoelectric and piezomagnetic cylinders deeming the contribution of the ultrathin and compliant elastic layer negligible, in this case. Otherwise, the elastic layer, being more complaint than the other active constituents (see material properties in Table 2), affects the effective stiffness of the composite structure, which in turn shifts the resonance frequency. It is important to note, as discussed earlier, the mechanical boundary conditions also affected the latter. Not only the frequency can be veered depending on the geometrical and material attributes of the elastic bonding layer, but the bandwidth can also be alternated by changing the thickness of this passive mediation layer, as discussed recently by Youssef et al. [56]. In all, these findings collectively point towards the suitability of the investigated composite structure for the development of tunable magnetic filters with importance in the high-frequency communication realm.

*3.3 Effect of the Demagnetization Field*

Figure 2b signifies the dependence of the demagnetization factor on the wall thickness of the piezomagnetic cylinder, where a decrease in the wall-thickness resulted in an increase in the effect of the demagnetization factor. On the contrary, an increase in the thickness showed that the effective magnetic field is nearly equivalent to the applied field, i.e., the demagnetization factor approaches zero. Keeping this in mind, the values of the direct magnetoelectric coupling coefficient were found to be influenced by the demagnetization effect that resulted in reducing the resulting DME, as expected, since the effective magnetic field was lowered by the demagnetization factor. On the other hand, also foreseen, the location of the resonant frequency is independent of the inclusion of the demagnetization effect in the calculations leading to the DME response since the essence of the demagnetization factor being a geometrical construct with no physical link to the mechanical, electrical, or magnetic properties of the constituents. To better illustrate the dependence, the difference between the DME coupling coefficients was calculated and plotted at the bottom of Figure 3 for each DME-frequency response amounting for a difference ranging between 7.35 to 25.53 $(V/m)(A/m)^{-1}$ with a strong dependence on the boundary conditions, as discussed before.

The DME of the composite cylinder at the first resonant frequency was reported to change due to the demagnetizing effect, from 19.1 to 3.1, 22.65 to 6.9, 29.6 to 4.1, and 10.2 to 2.9 $(V/m)(A/m)^{-1}$, respectively for the C-C, F-F, F-C, and C-F mechanical boundary conditions. This amounted to a decrease in the DME by 83.9%, 69.6%, 86.3%, and 72.1%, respectively. As discussed above, the effect of demagnetization energy is inevitable given its lineage to the geometry and size of the device under investigation. That is, while the results presented thus far compares the effect of the demagnetization on the DME response to elucidate its negative influence, it is imperative that the demagnetization factor must be included in future analytical modeling of the concentric cylinders. It is also important to note that the adverse influence of the demagnetization is amplified near the peripheries, in our case the inner and outer diameter, where the demagnetization factor is maximum and approaches nearly the same value as shown in Figure 2b. This is of specific importance to the spatial distribution of the radial strains that will degrade the localized, and in turn the global, magnetoelectric response of the composite structure.

*3.4 Effect of the Geometry*

Based on the preceding discussions, there are three overarching conclusions. First, the mechanical boundary conditions can be used to enhance the magnetoelectric response as a function of frequency by effectively changing not only the amplitude but also the resonant frequency. Second, the elastic layer, which is important for the practical assembly of the cylinder, also plays a notable role in the value and the resonant frequency of DME response. Finally, the demagnetization effect is imperative to account for given its major contribution to the over magnetic energy based on the geometry. However, these conclusions were curated based on specific geometrical attributes of the composite cylinder, leaving a gap in formalizing the overall dependence on geometry. It is then the objective of this section to demonstrate the interrelationship between the geometry and DME for all mechanical boundary conditions, in the presence of the bonding elastic layer, and by accounting for the demagnetization factor. In essence, Figure 4 presents a search schema within the design envelope to assist in identify the conditions leading to the maximum DME response.

Figure 4 plots the DME coupling coefficient as a function of two normalized geometrical parameters, namely the sensor phase ratio ($m = h_p/t_{total}$) and the inner radius ratio ($R = a/t_{total}$), at the corresponding first resonant frequency for each mechanical boundary condition. In the calculations leading to Figure 4, the effect of the elastic layer was also included by continuing to take the thickness to be 7.5 μm. The *m* ratio was taken to be [0.1:0.9], signifying that the piezoelectric cylinder occupying 10% to 90% of the overall thickness. The *R* ratio was [1:10] representing a range of inner radius that ranges from 5 mm to 50 mm. As evident from the previous results and the underlying elastodynamic response, the resonant frequency is dependent on the boundary condition and was calculated *a priori* for C-C, C-F, F-C, and F-F scenarios using

Equation B-3 included in the appendix. The derivation of the resonant frequency equations can be found in Appendix B.

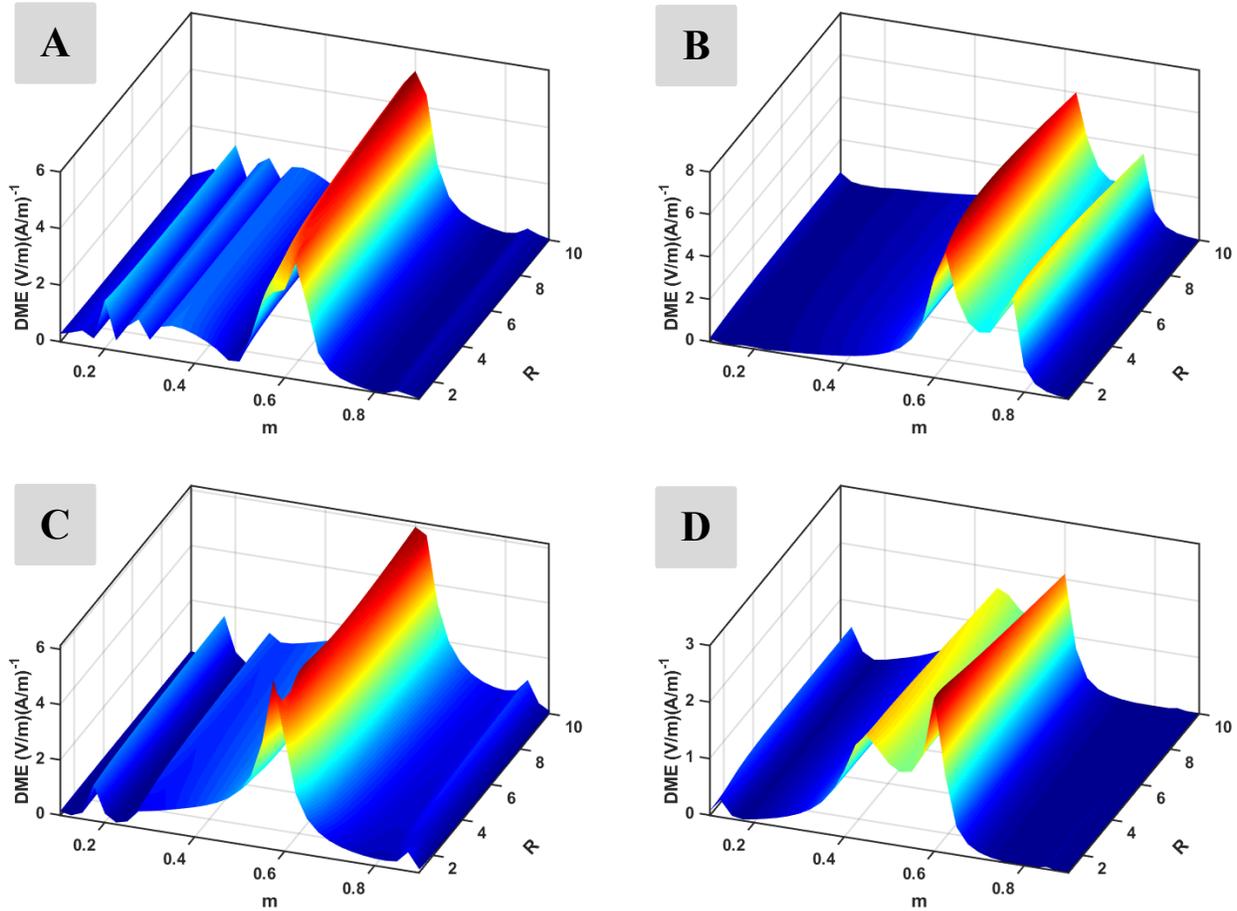

**Figure 4: The peak DME values at the resonant frequency as function of m and R ratio for (A) C-C, (B) F-F, (C) F-C, and (D) C-F boundary conditions.**

The maximum magnetoelectric coupling coefficient shows a higher dependency on the sensor phase ratio than the inner radius ratio, regardless of the boundary condition. That is, for a given $m$ value, the DME remain constant over the entire range of $R$ values, which is consistent with the previous results. The change in the geometry has a far reaching influence on the DME values specially when considering the geometry-driven demagnetization factor. The latter is dependent on the portion of the overall thickness associated with the inner piezomagnetic cylinder such that a change in *m* influences the fraction of Terfenol-D in the overall composite. In other words, as the *m* ratio increases, the thickness of the piezoelectric cylinder also increases resulting in a proportional reduction in the thickness of the piezomagnetic cylinder since the overall thickness was kept constant. This, in turn, amplifies the demagnetization factor and negatively affects the DME. On the other hand, the change in the inner radius, i.e., change in the *R* ratio, has no bearing on the demagnetization factor resulting in a constant DME over the entire range of R. In closing, Figure 4 and the associated equations and discussion completes the analytical modeling framework to fully investigate the considered concentric composite cylinder for direct magnetoelectric coupling with possible applications in tunable magnetic filters and energy harvesting.

On the limitation of the present model, it is worthy to note that while the current formulation includes several physical phenomena in regard to the dynamic response of concentric composite multiferroic cylinders, it includes two limitations. First, the model makes no attempt to account for the behaviors of the magnetic spins, electromagnetic waves, and acoustic waves, which require the amendment of the current framework to include the Landau-Lifshitz-Gilbert equation and Maxwell equations. Second, the model does not take into account the time and temperature dependent properties of the bonding elastic layer.

## 4. Conclusions

The presented model pursued the study of the geometrical effects on the direct magnetoelectric coupling response of concentric composite multiferroic cylinders. Concurrently investigated were the influence of the mechanical boundary and continuity conditions. The results obtained from the proposed analysis indicate that the shear lag mechanism has no bearing on the induced radial displacements in the composite cylinder, which in turn does influence the DME coefficient. On the other hand, the DME was found to show strong dependence on the geometrical-construct of the demagnetization effect, where the latter has a drastic, adverse effect on the calculated DME coefficient, whether the separating thin elastic layer was sanctioned or suppressed and regardless of the considered boundary conditions. Moreover, the harmonic frequencies at which the peak DME occurred also were proved to be independent of the demagnetization factor given it is a geometrical construct that has no effect on the stiffness or inertia of the constituents. Finally, the effect of the geometrical attributes of the composite cylinder were probed in details to elucidate the overall design space of these composites. While future research is warranted on the proposed model to include the dynamics of the multiferroic systems, the results are promising for future investigations in development of multiferroic-based devices such as magnetic filters and energy harvesters.

Appendix A

Considering Figure A-1 and according to the derivation of Joseph et al [59],

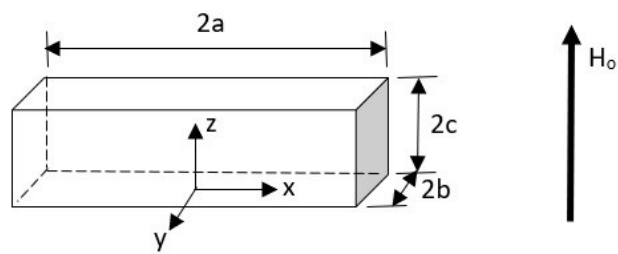

Figure 1-A Rectangular prism subjected to magnetic field H₀

The demagnetization factor in z-direction is,

$$N_{zz} = \left(\frac{1}{4\pi}\right)[cot^{-1}f(x,y,z) + cot^{-1}f(-x,y,z) + cot^{-1}f(x,-y,z) + cot^{-1}f(x,y,-z) + cot^{-1}f(-x,-y,z) + cot^{-1}f(x,-y,-z) + cot^{-1}f(-x,y,-z) + cot^{-1}f(-x,-y,-z)] \quad (A-1)$$

where,

$$f(x,y,z) = \frac{[(a-x)^2+(b-y)^2+(c-z)^2]^{\frac{1}{2}}(c-z)}{(a-x)(b-y)} \quad (A-2)$$

Assuming that it is required to obtain the demagnetization factor in the (x-z) plane, at which y=0, hence,

$$N_{zz} = \left(\frac{1}{4\pi}\right)[2cot^{-1}f(x,z) + 2cot^{-1}f(-x,z) + 2cot^{-1}f(x,-z) + 2cot^{-1}f(-x,-z)] \quad (A-3)$$

and,

$$f(x,z) = \frac{[(a-x)^2+(b)^2+(c-z)^2]^{\frac{1}{2}}(c-z)}{(a-x)(b)} \quad (A-4)$$

Converting to polar coordinates and assuming that,

$$x = r\theta \qquad z = r \qquad 2b = w \qquad 2c = h \qquad 2a = 2\pi r_m$$

substituting in Equations (A-3) and (A-4), to obtain,

$$N_{zz} = \frac{1}{4\pi}[2cot^{-1}f(r\theta,r) + 2cot^{-1}f(-r\theta,r) + 2cot^{-1}f(r\theta,-r) + 2cot^{-1}f(-r\theta,-r)]$$

(A-5)

where,

$$f(r\theta,r) = \frac{\left[(\pi r_m-r\theta)^2+\left(\frac{w}{2}\right)^2+\left(\frac{h}{2}-r\right)^2\right]^{\frac{1}{2}}\left(\frac{h}{2}-r\right)}{(\pi r_m-\theta r)\left(\frac{w}{2}\right)} \quad (A-6)$$

**Appendix B**

In order to investigate the harmonic frequencies of the system, the composite cylinder is modeled as shown in Figure B-1 using lumped parameters approach.

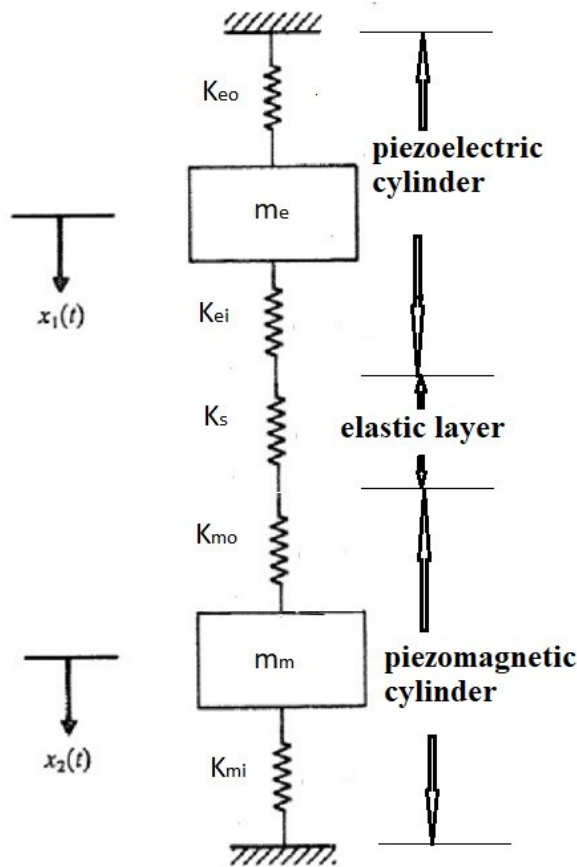

**Figure B-1: Dynamic model of the composite cylinder**

The equation of motion for the above model can be written in matrix form as

$$\begin{bmatrix} m_e \frac{d^2}{dt^2} + K_{eo} + K_e & -K_e \\ -K_e & m_m \frac{d^2}{dt^2} + K_{mi} + K_e \end{bmatrix} \begin{bmatrix} x_1 \\ x_2 \end{bmatrix} = \begin{bmatrix} 0 \\ 0 \end{bmatrix} \quad \text{(B-1)}$$

where, $K_e$ is the equivalent stiffness, found to be,

$$K_e = \frac{K_{ei} K_s K_{mo}}{K_{ei} K_s + K_s K_{mo} + K_{mo} K_{ei}} \quad \text{(B-2)}$$

Equation A-2 was then solved to arrive to the resonant frequency equation

$$\omega_1^2, \omega_2^2 = \frac{1}{2}\left\{\frac{(K_{eo} + K_e)m_m + (K_e + K_{mi})m_e}{m_e m_m}\right\}$$

$$\pm \frac{1}{2}\left[\left\{\frac{(K_{eo} + K_e)m_m + (K_e + K_{mi})m_e}{m_e m_m}\right\}^2\right.$$

$$\left. - 4\left\{\frac{(K_{eo} + K_e)(K_e + K_{mi}) - K_e^2}{m_e m_m}\right\}\right]^{\frac{1}{2}} \qquad (B-3)$$

where, $m_e$ is the mass of the piezoelectric cylinder ($m_e = \rho_e \pi(c^2 - b^2)$) and $m_m$ is the mass of the piezomagnetic cylinder ($m_m = \rho_m \pi(b^2 - a^2)$) while taken $\rho_e$ as the mass density of the piezoelectric material and $\rho_m$ as the mass density of the piezomagnetic material. The stiffness of each layer is evaluated by

$$K = \frac{2\pi r_{mean} E(1-v)}{t_{thickness}(1+v)(1-2v)} \qquad (B-4)$$

where, E and v are the modulus of elasticity and Poisson's ratio of the material respectively. For the C-C boundary condition all stiffness have values, while for the F-F boundary condition ($K_{eo}=K_{mi}=0$), the F-C boundary condition ($K_{mi}=0$) and for the C-F boundary condition ($K_{eo}=0$).